# Maternal education and offspring birthweight for gestational age: the mediating effect of smoking during pregnancy

Aurélie Nakamura[a,b,*], Laura Pryor[a,c], Morgane Ballon[b,d], Sandrine Lioret[d], Barbara Heude[d], Marie-Aline Charles[d], Maria Melchior[a], Fabienne El-Khoury Lesueur[a] a Sorbonne Université, INSERM, Institut Pierre Louis d'Épidémiologie et de Santé Publique (IPLESP), Department of Social Epidemiology, Paris, France
b French School of Public Health (EHESP), Doctoral Network, Rennes, France
c Johns Hopkins Bloomberg School of Public Health, Department of International Health, Baltimore, Maryland, USA
d INSERM, UMR1153 Epidemiology and Biostatistics Sorbonne Paris Cité Center (CRESS), Early Origin of the Child's Health and Development Team (ORCHAD), Paris Descartes University, Paris, France

\* Corresponding author:

Aurélie Nakamura
Sorbonne Université, INSERM, Pierre Louis Institute of Epidemiology and Public Health, Department of Social Epidemiology
27 rue Chaligny 75012 Paris, France
aurelie.nakamura@iplesp.upmc.fr
+33681817496 - fax +331 44 73 84 62

**Abstract**

**Background** Small for gestational age (SGA) birthweight, a risk factor of infant mortality and delayed child development, is associated with maternal educational attainment. Maternal tobacco smoking during pregnancy could contribute to this association. We aimed to quantify the contribution of maternal smoking during pregnancy to social inequalities in child birthweight for gestational age (GA).

**Methods** Data come from the French nation-wide ELFE cohort study, which included 17,155 singletons. Birthweights for GA were calculated using z-scores. Associations between maternal educational attainment, tobacco smoking during pregnancy and child birthweight for GA were ascertained using mediation analysis. Mediation analyses were also stratified by maternal pre-pregnancy body mass index.

**Results** Low maternal educational attainment was associated with an increased odd of tobacco smoking during pregnancy (adjusted OR (ORa)=2.58 [95% CI 2.34, 2.84]) as well as a decrease in child birthweight for GA (RRa=0.94 [95% 0.91, 0.98]). Tobacco smoking during pregnancy was associated with a decrease in offspring birthweight for GA (RRa=0.73 [95% CI 0.70, 0.76]). Mediation analysis suggests that 39% of the effect of low maternal educational attainment on offspring birthweight for GA was mediated by smoking during pregnancy. A more important direct effect of maternal educational attainment on child birthweight for GA was observed among underweight women (RRa=0.82 [95%CI 0.72, 0.93]).

**Conclusions** The relationship between maternal educational attainment and child birthweight for GA is strongly mediated by smoking during pregnancy. Reducing maternal smoking could lessen the occurrence of infant SGA and decrease socioeconomic inequalities in birthweight for GA.

## Introduction

Approximately 5 to 15% of singletons born in Western countries were born small for gestational age (SGA), which is a risk factor for mortality,[1,2] delayed development and poor health throughout the lifecourse.[3] Risk factors for SGA birth include socioeconomic disadvantage,[4] possibly because of differences in health behaviors,[5] health care access and health literacy across socioeconomic groups.[6] Women's socioeconomic position can be assessed by the highest level of education attained, considered one of the best predictors of pregnancy outcomes.[7]

One of the health behaviours most strongly associated with maternal educational attainment is tobacco smoking,[8] which is also an important risk factor for adverse perinatal outcomes.[9] Prior research suggests that maternal smoking early in pregnancy contributes to social inequalities in SGA,[10–12] however its role throughout pregnancy has not been thoroughly documented. In addition, prior studies neither systematically accounted for important confounders nor applied rigorous methods to study mediation. One study based on multiple mediation analyses[13] and one study based on path analysis[14] also suggested smoking during pregnancy could be a mediator in the relation between maternal educational attainment and SGA offsprings. In France, women's tobacco smoking rates are among the highest in Europe: approximately 30% of women of child bearing age are daily smokers,[14] and 17% use tobacco throughout pregnancy.[15] Socioeconomically disadvantaged women have higher levels of smoking and are more likely to continue smoking through pregnancy than those who have more favourable backgrounds.[16] At the same time, low maternal educational attainment is associated with a higher maternal body mass index (BMI),[17] which could be protective against having a SGA offspring,[18] therefore maternal pre-pregnancy BMI needs to be accounted for when examining the contribution of tobacco smoking.[19]

In this study we quantify the mediating effect of tobacco smoking during pregnancy in the association between maternal educational attainment and offspring birthweight for gestational age (GA) in a large French nation-wide birth cohort study (ELFE).

## Methods

*Study design and participants*

ELFE (*Etude Longitudinale Française depuis l'Enfance*) is a nationally-representative birth cohort that follows 18,275 children born in France in 2011.[20] Children were recruited in 320 maternity wards using random sampling. To be included, children had to be singletons or twins born at 33 weeks of gestation or older. Mothers had to be at least 18 years of age and not planning to move outside of Metropolitan France in the three years following study inclusion. Mothers had to be able to give consent in French, English, Arabic or Turkish. Data were collected at birth via face-to-face interviews conducted by midwives and via self-reported questionnaires. The participation rate was of 51%.[20] The present analysis included a sample of 17,155 singletons with complete data on birthweight, gestational age and sex. The ELFE study received the approval of France's bodies regulating ethical research conduct (Comité Consultatif sur le Traitement des Informations pour la Recherche en Santé: CCTIRS; Commission National Informatique et Libertés: CNIL).

*Measures*

*Maternal educational attainment*

Maternal educational attainment was categorized into 1) high (>high school degree) or 2) low (≤high school degree).

*Tobacco smoking during pregnancy*

Tobacco smoking during pregnancy was assessed retrospectively in face-to-face interviews conducted at the time of the child's birth: 'Did you smoke during your pregnancy, even if it was from time to time?' (yes/no).

*Child birthweight for gestational age*

The child's birthweight was extracted from medical records. Birthweights were normalized according to GA and sex (z-scores). [21]

*Covariates*

Several factors potentially associated with either smoking during pregnancy[16] or child birthweight[22] were examined, including demographic characteristics: maternal age at birth, nationality (French vs. other), parity (0 vs. ≥1), residential area (Paris region vs. other), relationship status (living with a partner: yes vs. no); maternal health and perinatal characteristics: body mass index (BMI) before pregnancy (in kg/m$^2$), attendance of at least one antenatal education class (yes vs. no), number of prenatal visits, severe psychological difficulties during pregnancy (yes vs. no), alcohol consumption during pregnancy (>once a month vs. ≤once a month), pregnancy complications (e.g. high blood pressure, gestational diabetes, bleeding, etc.: ≥1 vs. <1 complication), gestational weight gain (in kg), physical activity during pregnancy (based on the Pregnancy Physical Activity Questionnaire [23] where the overall score, ranging from 0.2 to 1292.4, was categorized as lower vs. greater than the median value (158.9, mean=175.6 (88.9)) or missing: physically active vs. physically inactive vs. missing). We also controlled for maternity type (public vs. private) and child sex.

Birthweight, GA, child sex and pregnancy complication information was obtained through medical records. All other variables were self-reported by face to face interview, except for physical activity during pregnancy that was reported by telephone interview. [20]

*Statistical analyses*

First, bivariate analyses were conducted to test associations between 1) maternal educational attainment and tobacco smoking during pregnancy, 2) maternal educational attainment and offspring birthweight for GA and 3) maternal tobacco smoking during pregnancy and offspring birthweight for GA. Then, using multivariable linear regression models, we adjusted for covariates associated either with maternal educational attainment, tobacco smoking during pregnancy or offspring birthweight for GA in bivariate analyses.

Then, we analysed associations between maternal educational attainment, tobacco smoking during pregnancy and offspring birthweight for GA in a counterfactual mediation analysis model, as defined by VanderWeele and colleagues[24]. In brief, this method allows the decomposition of the effects of maternal educational attainment on offspring birthweight for GA in a natural direct effect and an indirect effect passing through maternal tobacco smoking during pregnancy in a single model. Direct and indirect effects of maternal educational attainment and the proportion of the association with offspring birthweight mediated by tobacco smoking during pregnancy were estimated as division of the indirect effect on the total effect[24]. Figure 1 illustrates the theoretical model in a diagram acyclic graph (DAG) of the hypothesized mediation mechanism. In additional analyses, we tested for an interaction between maternal educational attainment (exposure) and tobacco smoking during pregnancy (potential mediator). The mediation analysis was carried out on complete observations.

Finally sensitivity analyses regarding robustness of the results to potential unmeasured or uncontrolled confounders were conducted by calculating E-values. The E-value "represents minimum strength of association, on the risk ratio scale, that an unmeasured confounder would need to have with both the treatment and the outcome to fully explain away a specific treatment-outcome association, conditional on the measured covariates".[25] The closer to one the E-value is, the greater are chances that unmeasured confounders could bias our estimates. In our mediation model, unmeasured confounders could affect estimates of associations between 1) maternal education attainment and maternal tobacco smoking during pregnancy 2) maternal tobacco smoking and offspring birthweight for GA and 3) maternal education attainment and offspring birthweight for GA. Thus, three E-values were calculated.

All analyses were conducted with SAS 9.4. (SAS Inst Cary NC. 2003) Mediation analyses were implemented using the SAS macro "%mediation" developed by Valeri and VanderWeele [26]. E-values and 95% confidence intervals were calculated using the R package EValue.[27]

In supplementary analyses, to account for the possible moderating role of BMI in the relationship between mother's education level and child's birthweight, we repeated the mediation analyses stratifying by pre-pregnancy maternal BMI, categorized as 1) low (<18.5kg/m$^2$), 2) normal (≥18.5 and <25kg/m$^2$), 3) or overweight (≥25 and <30kg/m$^2$) and high (≥30kg/m$^2$).

## Results

*Characteristics of ELFE cohort study participants*

Descriptive statistics of women included in the analysis are presented in Supplementary Table 1S. About 60% of mothers had high educational attainment. Twenty percent of mothers reported smoking tobacco during pregnancy. Children's birthweight was on average 3,335g (479 g). The percentage of missing data varied between 0 and 3%, except for physical activity during pregnancy (25% missing values). Compared to children included in the study, those who were excluded from the analyses because of missing values on covariates (n=2,737) were significantly more likely to be born to mothers who had low educational attainment (44% vs. 41%, *P*<0.001), but there were no differences in terms of pre-term births (5% vs. 4%, P=0.1), birthweight (3,330g (507g) vs. 3,335g (474g), P=0.6) and low birthweight for GA (9% vs. 8%, *P*=0.4) (data available upon request).

*Mediation analysis*

Women with low educational attainment had an increased risk of having a child with decreased birthweight for GA compared to those who had high educational attainment (crude RR=0.94 [95% CI 0.91, 0.97]). After adjustment for potential confounders (all covariates listed above), this association remained stable (RRa=0.94 [95% CI 0.91, 0.98]). Low educational attainment was also associated with an increased odd of tobacco smoking during pregnancy, even after adjustment for potential confounders (ORa=2.58 [95% CI 2.35, 2.84]). Tobacco smoking during pregnancy was associated with a decreased birthweight for GA, even after adjustment for potential confounders (RRa=0.73 [95% CI 0.70, 0.76) (Figure 2).

Women's educational attainment was associated with tobacco smoking during pregnancy, which in turn was associated with offspring birthweight for GA, allowing for mediation analyses. Natural direct and indirect effects and proportion mediated of birthweight weight for GA, without interactions between educational attainment and tobacco smoking during pregnancy, are presented in Supplementary Figure 1S. We found a natural direct effect of educational attainment on offspring birthweight for GA ($RR^{direct}$=0.94 [95% CI 0.91, 0.98]). We also observed a statistically significant adverse natural indirect effect of low educational attainment on offspring birthweight for GA via an increase in tobacco smoking during pregnancy ($RR^{indirect}$=0.96 [95% CI 0.95, 0.96]). In additional analyses, we included an interaction between educational attainment and tobacco smoking during pregnancy in our models, yielding comparable direct and indirect risk ratios. The interaction term between educational attainment and tobacco smoking during pregnancy was not statistically significant (*P*= 0.6), thus we did not include it in further analyses. Overall, we estimated that 39% of the effect of maternal educational attainment on offspring birthweight for GA was mediated by maternal tobacco smoking during pregnancy.

E-values were estimated at 1.32 [95%CI 1.16, +inf[ for adjusted maternal educational attainment and offspring birthweight for gestational age association; 2.08 [95% CI 1.96, +inf[ for maternal tobacco smoking and offspring birthweight for GA and 2.59 [95%CI 2.44, + inf[ for maternal educational attainment and maternal tobacco smoking associations.

*Stratification by maternal pre-pregnancy BMI*

The association between maternal educational attainment and offspring birthweight for GA was stronger in underweight women (RRa=0.82 [95% CI 0.72, 0.93]) than in women with normal or high BMI. The association between maternal tobacco smoking during pregnancy and infant birthweight for gestational age was stronger in overweight women (RRa=0.70 [95% CI 0.63, 0.77]) than in other groups. (Table 1)

Nevertheless, the indirect effect of maternal educational attainment on infant birthweight for GA, mediated by tobacco smoking during pregnancy, was consistent across all strata of maternal BMI. Thus, the proportion of the effect of maternal educational attainment mediated (PM) by tobacco smoking was lower in underweight women (PM=23%) than in women with normal BMI (PM=47%). (Figure 3)

## Discussion

*Principal findings*

Studying a large national cohort of women and their children, we found that low maternal educational attainment predicts a 1.1 fold decrease in child birthweight for GA. Up to 39% of this association appears to be due to differences in maternal tobacco smoking during pregnancy. Importantly, tobacco smoking plays a lesser role with regard to social inequalities in birthweight for GA in children of women who are underweight and who might have other specific risk factors. To the extent that birthweight predicts a number of health outcomes later in life (e.g. psychological and cognitive development, overweight/obesity, cardiovascular disease),[28,29] exposure to tobacco smoking in pregnancy could play an important role in socioeconomic disparities in health throughout the life course.

Our findings are in line with prior studies conducted in other settings and using other methods. In a study conducted in the Netherlands, van den Berg and colleagues[10,30] found that tobacco smoking during pregnancy mediated both the associations between maternal educational level and offspring low birth weight and maternal educational attainment and SGA births, as calculated by the percentage variation of the odds ratio due to smoking during pregnancy. In a study conducted in France, Ballon and colleagues [13] also found that smoking during pregnancy was a significant mediator of the effects of low educational attainment in low birthweight and small offspring. Using structural equation models, Da Silva and colleagues found that maternal socioeconomic position indirectly influenced child birthweight through maternal and paternal smoking during pregnancy in Brazil. [31] Therefore our data are largely consistent with research conducted in other settings, which did not apply the same rigorous method to test causal inference as we did.

*Potential underlying mechanisms of smoking during pregnancy in the effects of maternal educational attainment on offspring birthweight for GA*

This study contributes to the growing body of literature supporting the hypothesis of embodiment of social inequalities. Women with a more disadvantaged educational attainment are more likely to smoke tobacco during pregnancy, [8] and less likely to quit smoking.[32] In a systematic review and a small meta-analysis, financial incitation on disadvantaged women has shown a potential positive effect on smoking cessation.[33]

Our results also suggest an important effect of smoking during pregnancy on reducing offspring birthweight for GA, even after adjusting for a large number of potential confounders, including maternal educational attainment. This is in accordance with a substantial body of scientific literature reporting an association between maternal tobacco smoking and offspring birthweight for GA.[9] Nicotine crosses the placenta and foetal nicotine concentrations could be increased by 15% of maternal (blood) concentrations, and could affect foetal growth and birthweight.[34] The effect of maternal tobacco smoking during pregnancy on birthweight for gestational age could also reflect gene-environment interactions between tobacco use and genes that characterize the nicotine receptor and are associated with nicotine dependence. Women with low educational attainment are also more likely to face stress events during pregnancy such as being single mothers or having shift work, which increase their probability of not quitting smoking during pregnancy and of having smaller babies. [36]

*Strengths of the study*

In terms of strengths, it is worth pointing out that our analyses are based on a large birth cohort study where children's birthweights were ascertained at the maternity ward and documented in medical records and controlled for a large number of potential confounding variables. Moreover, birthweights were assessed using z-scores which take into account GA and sex of the offspring. In addition, the assessment of the associations between maternal educational attainment, tobacco smoking during pregnancy and offspring birthweight for GA remained stable after adjustment for potential confounders, despite the important number of missing values. Finally, the use of mediation models, particularly well-suited for the study of mechanisms underlying socioeconomic inequalities in health, is novel.

*Limitations of the data*

Our study has a number of limitations which should be considered. First, maternal educational attainment has been used as a proxy of socioeconomic position (SEP) even though SEP is a complex construct including other dimensions then education. However, maternal educational attainment have been shown as a good proxy to study the effect of SEP on children's health[37] as most financial resources result from salaries, thus from work, which access is often conditioned by educational attainment.[37] Then, ELFE is a cohort study based on voluntary participation. As a result the proportion of mothers with a low education level is lower than in the general population of France. Maternal smoking during pregnancy was self-reported at birth, which could lead to recall and desirability bias. However, the prevalence of maternal smoking in pregnancy in ELFE (approximately 20%) is consistent with national figures.[15] Our measure of smoking during pregnancy did not take into account the frequency or the quantity of tobacco smoker. Sensitivity analyses on self-reported tobacco smoking in the third trimester of pregnancy (17% of the sample) lead to similar results as the ones regarding the whole pregnancy (data not shown) but women who smoked only in early pregnancy might have declared they did not smoke during pregnancy. We may have underestimated social inequalities in birthweight for GA but our results are generalizable. Second, our analyses were based on complete cases, resulting in the exclusion of 5,500 participants. Mothers who were excluded due to missing values were younger, had lower educational attainment and were more likely to be non-French citizens. The direction of bias resulting from this exclusion is not obvious. Of note, only children born at 33 weeks of gestation or above were recruited, thereby limiting the results to babies not born severely premature. Severe prematurity[34] have other determinants and it is not clear if the contribution of tobacco smoking to social inequalities in this area would be similar as we observed. We thus acknowledge that the contribution of tobacco smoking to socioeconomic inequalities in infant birthweight should be examined in studies that include children born at younger gestational ages. Third, in France, levels of maternal tobacco smoking in pregnancy are relatively high (20% prevalence), but comparable to several other settings including Denmark and the Netherlands and some US States (e.g. West Virginia, Kentucky, Montana, Vermont, Missouri), which also calls into question the generalizability of our findings [39,40]. Fourth, a simple mediation analysis was conducted, which might conceal the contribution of other potential mediators. Nevertheless, it is unlikely that these other mediators (e.g. maternal nutrition), would confound the role of tobacco smoking. Tobacco smoking during pregnancy was self-reported by the mother and could be underestimated, even though our estimations of tobacco smoking during pregnancy are similar to those obtained in other French studies.[15,28]

Finally, sensitivity analyses pointed out some limitation to any causal interpretation of the observed interpretation between maternal educational attainment, maternal tobacco smoking and offspring birthweight for GA. Especially, we found a E-value of 1.3 [95%CI 1.2, +inf[ for the association between maternal educational attainment and offspring birthweight for GA, meaning any unmeasured factor being associated with a strength of a RR greater or equal to 1.4 could explain away the observed association between these two variables, after adjustment for measured potential confounders. Such factor could be any factor related to maternal diet during pregnancy for example.

Our study shows that the effects of maternal educational attainment on offspring birthweight for GA are in large part (39%) mediated by tobacco smoking during pregnancy. These data contribute to existing evidence that reducing maternal tobacco smoking during pregnancy could improve offspring as well as maternal perinatal outcomes, including birthweight for GA, and contribute to reducing socioeconomic inequalities in children's perinatal health. Interventions encouraging female smokers – particularly those who belong to socioeconomically disadvantaged groups – to quit smoking prior to or during pregnancy should be a public health priority across industrialized countries. However, more specific psychosocial interventions targeting women who are more at risk of smoking during pregnancy and having children with low birthweight for GA should be developed. These interventions should take into consideration women's socioeconomic and psychosocial characteristics, such as their ability to access social workers for improving their living conditions and to get support from their own network and health professionals during their pregnancy and after child birth.

**Funding**

The Elfe survey is a joint project between the French Institute for Demographic Studies (INED) and the National Institute of Health and Medical Research (INSERM), in partnership with the French blood transfusion service (Etablissement français du sang, EFS), Santé Publique France, the National Institute for Statistics and Economic Studies (INSEE), the Direction générale de la santé (DGS, part of the Ministry of Health and Social Affairs), the Direction générale de la prévention des risques (DGPR, Ministry for the Environment), the Direction de la recherche, des études, de l'évaluation et des statistiques (DREES, Ministry of Health and Social Affairs), the Département des études, de la prospective et des statistiques (DEPS, Ministry of Culture), and the Caisse nationale des allocations familiales (CNAF), with the support of the Ministry of Higher Education and Research and the Institut national de la jeunesse et de l'éducation populaire (INJEP). Via the RECONAI platform, it receives a government grant managed by the National Research Agency under the "Investissements d'avenir" programme (ANR-11-EQPX-0038).

Additionally, this analysis was funded by the SOCIALRISK_MH project funded through the ANR 'Social determinants of health' program (2012), where Maria Melchior is also the Principal Investigator.

**Conflicts of interest**

The authors declare that they have no competing interests

**Key points**

- 39% of social inequalities of infant birthweight for gestational age were mediated by maternal tobacco smoking during pregnancy
- Direct effect of maternal educational attainment on offspring birthweight for gestational age was stronger in underweight women
- Preventing maternal smoking during pregnancy, especially in socially disadvantaged women, could reduce social inequalities in children's health

**Table 1**– Adjusted risk ratios (RR) and odds ratios (OR) for the association between 1) offspring birthweight for gestational age and maternal educational attainment 2) offspring birthweight for gestational age and smoking during pregnancy 3) smoking during pregnancy and maternal educational attainment stratified by body mass Index (BMI) in four categories (<18.5; 18.5-<25;25-<30; ≥ 30kg/m$^2$), ELFE Study, France, 2011

| Exposure/Outcome | Offspring birth weight for gestational age Adjusted RR [95% CI] | Maternal smoking during pregnancy Adjusted OR [95% CI] |
|---|---|---|
| *BMI < 18.5kg/m2, n=1,117* | | |
| Maternal educational level | 0.82 [0.72, 0.93] | 3.17 [2.30, 4.38] |
| Maternal smoking during pregnancy | 0.78 [0.69, 0.88] | |
| *18.5 ≤ BMI < 25.0kg/m2, n=9,520* | | |
| Maternal educational level | 0.95 [0.91, 0.99] | 2.66 [2.56, 2.99] |
| Maternal smoking during pregnancy | 0.72 [0.69, 0.76] | |
| *25.0 ≤ BMI < 30.0kg/m2, n=2,495* | | |
| Maternal educational level | 0.96 [0.88, 1.04] | 2.09 [1.66, 2.65] |
| Maternal smoking during pregnancy | 0.70 [0.63, 0.77] | |
| *BMI ≥ 30.0kg/m2, n=1,406* | | |
| Maternal educational level | 0.95 [0.84, 1.07] | 2.27 [1.66, 3.10] |
| Maternal smoking during pregnancy | 0.77 [0.66, 0.87] | |

Adjustment for age at birth, nationality, parity, residential area, relationship status, attendance of at least one class to prepare for childbirth, number of prenatal visits, severe psychological difficulties during, alcohol consumption during pregnancy, pregnancy complications, gestational weight gain, physical activity during pregnancy, maternity type and child sex.

**Figure 1 –** Diagram acyclic graph (DAG), maternal educational attainment, maternal tobacco smoking during pregnancy, offspring's birthweight for gestational age and maternal prepregnancy body mass index

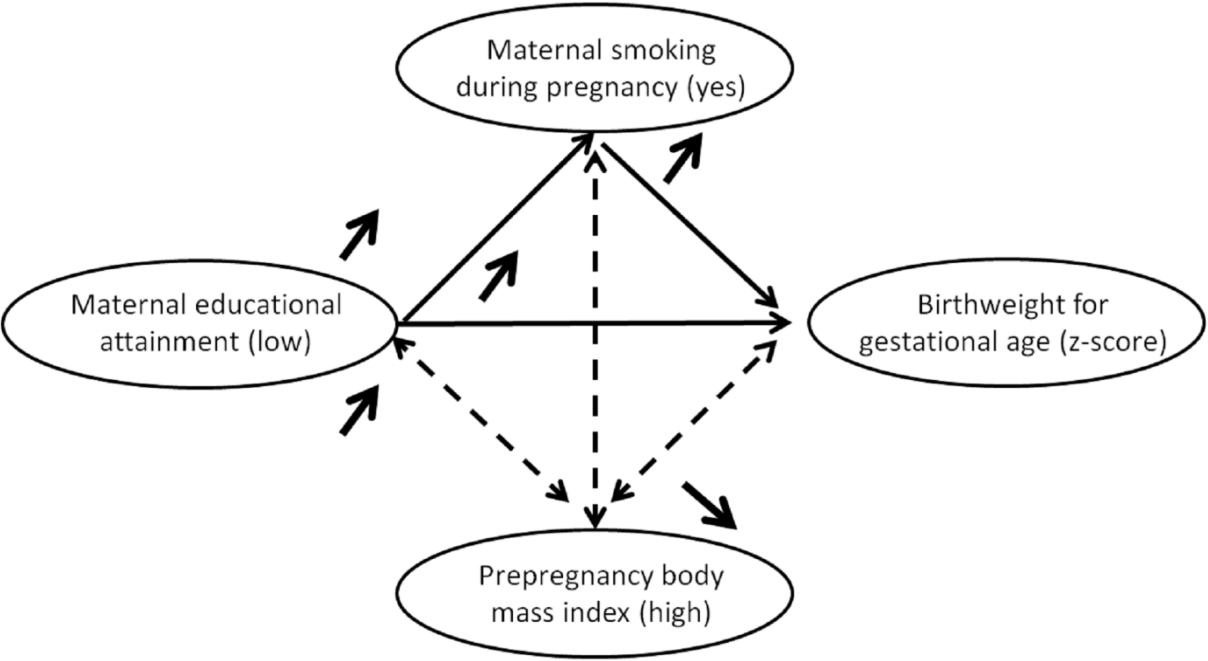

Full arrows correspond to causal relations and bidirectional dotted arrows correspond to non-causal correlations.

**Figure 2** – Crude and adjusted risk ratios and odds ratios for the association between 1) offspring birthweight for gestational age and maternal educational attainment 2) offspring birthweight for gestational age and smoking during pregnancy 3) smoking during pregnancy and maternal educational attainment (ELFE study, France, 2011)

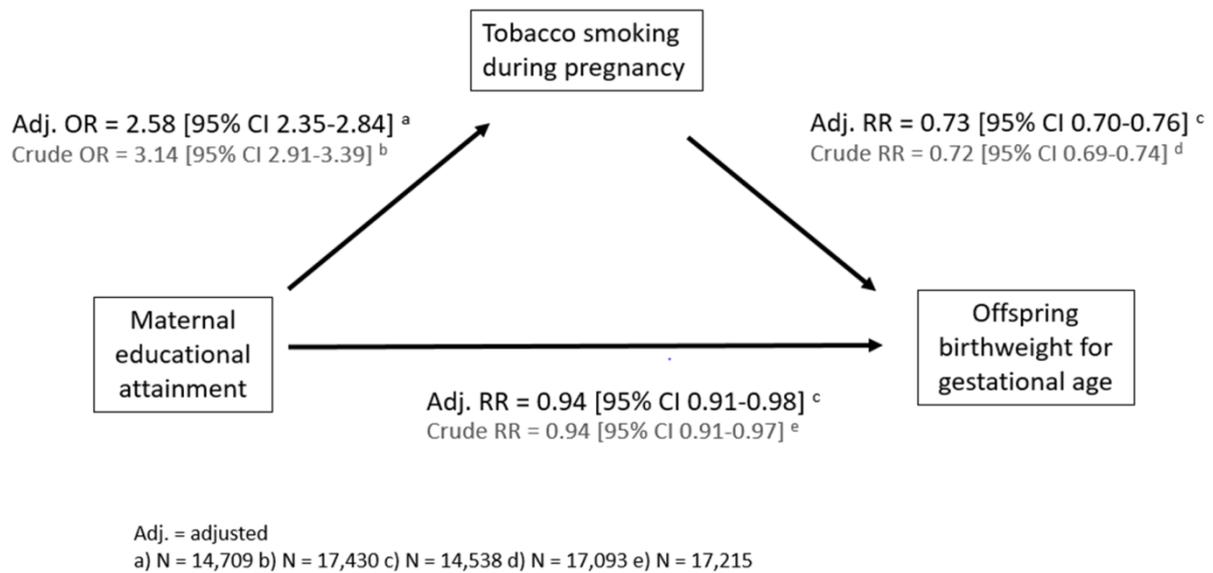

Adj. = adjusted

a) N = 14,709 b) N = 17,430 c) N = 14,538 d) N = 17,093 e) N = 17,215

Crude OR 1 was calculated on total sample size while crude OR 2 on multivariable complete case sample.

Adjustment for age at birth, nationality, parity, residential area, relationship status, body mass index (BMI) before pregnancy, attendance of at least one class to prepare for childbirth, number of prenatal visits, severe psychological difficulties during, alcohol consumption during pregnancy, pregnancy complications, gestational weight gain, physical activity during pregnancy, maternity type and child sex.

**Figure 3** - Mediation analysis between maternal educational attainment (exposure), maternal smoking during pregnancy (potential mediator) and offspring birthweight for gestational weight (outcome), A) for underweight women (BMI < 18.5kg/m$^2$, n=1,117) B) for women with a normal BMI (18.5 ≤ BMI < 25.0kg/m$^2$, n= 9,520) C) for overweight women (25.0 ≤ BMI < 30.0kg/m$^2$, n= 2,495) D) for obese women (BMI ≥ 30.0kg/m$^2$, n=1,406), ELFE French Study, 2011

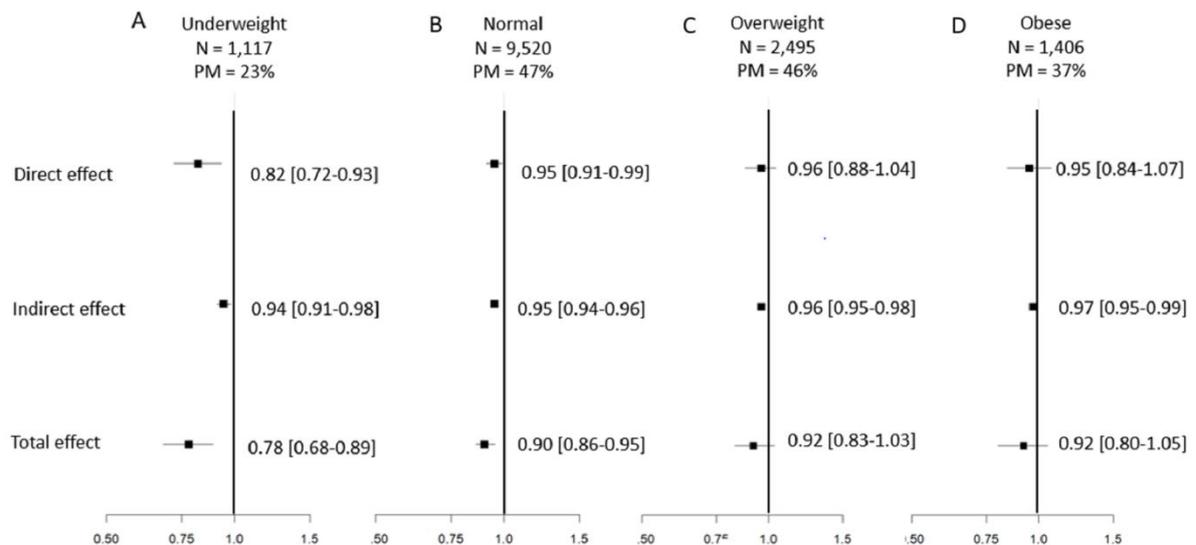

BMI = body mass index, PM = proportion mediated

Adjustment for age at birth, nationality, parity, residential area, relationship status, attendance of at least one class to prepare for childbirth, number of prenatal visits, severe psychological difficulties during, alcohol consumption during pregnancy, pregnancy complications, gestational weight gain, physical activity during pregnancy, maternity type and child sex.

**Supplementary material**

**Table 1S -** Characteristics of Women Participating in the ELFE Study, France, 2011, n=17,155 (n, %[a]/mean, sd)

| | n (%) or mean (sd) |
|---|---|
| *Maternal sociodemographic characteristics* | |
| Age at the child's birth | 30.6 (5.1) |
| French nationality (yes) | 15,608 (91.8%) |
| Parity (>0) | 9,171 (54.2%) |
| Residential area (Paris region) | 3,493 (20.4%) |
| Lives with a partner (yes) | 16,080 (94.5%) |
| | |
| *Maternal socioeconomic characteristics* | |
| Educational level (higher) | 10,276 (60.1%) |
| | |
| *Maternal health and perinatal characteristics* | |
| Pre-pregnancy BMI (kg/m$^2$) | |
|   <18.5 | 1,581 (9.2%) |
|   >=18.5 and <25 | 10,962 (63.9%) |
|   >=25 and <30 | 2,928 (17.1%) |
|   >=30 | 1,684 (9.8%) |
| Antenatal class (yes) | 9,368 (55.5%) |
| Number of prenatal visits | 8.5 (2.6) |
| Psychological difficulties during pregnancy (yes) | 2,127 (12.5%) |
| Smoking during pregnancy (yes) | 3,434 (20.2%) |
| Alcohol consumption during pregnancy (>1/month) | 1,240 (7.3%) |
| Complications during pregnancy (>=1) | 5,062 (29.5%) |
| Weight gain during pregnancy (kg) | 13.1 (5.6) |
| Type of maternity (public) | 11,287 (65.8%) |
| Physically active during pregnancy | |
|   No | 6,432 (37.5%) |
|   Yes | 6,540 (38.1%) |
|   Missing | 4,183 (24.4%) |
| *Child characteristics* | |
| Child gestational age (weeks) | 39.2 (1.4) |
| Pre-term birth (yes) | 403 (4.4%) |
| Birthweight (g) | 3,335.1 (479.2) |
| Small for gestational age (yes) | 1,446 (8.4%) |
| Sex (girl) | 8,294 (48.3%) |

**Figure 1S –** Mediation analysis between maternal education (exposure), smoking during pregnancy (potential mediator) and offspring birthweight for gestational age (outcome) (ELFE Study, France, n=14,538, 2011)

PM = proportion mediated

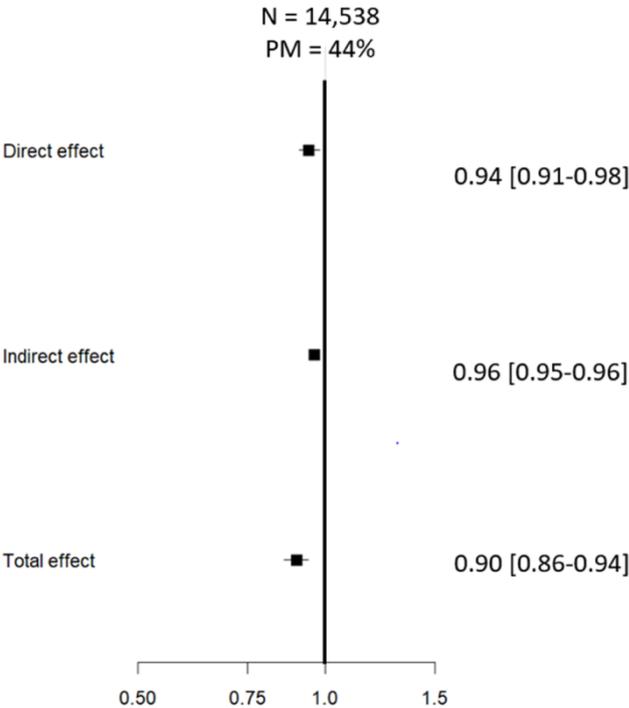

Adjustment for age at birth, nationality, parity, residential area, relationship status, body mass index (BMI) before pregnancy, attendance of at least one class to prepare for childbirth, number of prenatal visits, severe psychological difficulties during, alcohol consumption during pregnancy, pregnancy complications, gestational weight gain, physical activity during pregnancy, maternity type and child sex.